\documentclass[twocolumn, twocolappendix]{aastex631}
\usepackage{amsmath}

\begin{document}
\title{Signature of hadronic emission in gamma-ray spectrum of B2 1308+326}
\shorttitle{B2 1308+326: A very high energy FSRQ}
\shortauthors{A. A. Dar et. al}
\author[0009-0008-9681-8224]{Athar A. Dar}
\correspondingauthor{Athar A. Dar}
\email{ather.dar6@gmail.com}

\affiliation{Department
of Physics, University of Kashmir, Srinagar 190006, India}
\affiliation{Department of Physics, Central University of Kashmir, Ganderbal 191201, India}

\author{Zahir shah}
\affiliation{Department of Physics, Central University of Kashmir, Ganderbal 191201, India}
\correspondingauthor{Zahir Shah}
\email{shahzahir4@gmail.com}
\author{Sunder Sahayanathan}
\affiliation{Astrophysical Sciences Division, Bhabha Atomic Research Center, Mumbai 400085, India}
\affiliation{Homi Bhabha National Institute, Mumbai 400094, India} 
\email{sunder@barc.gov.in}
\author{Naseer Iqbal}
\affiliation{Department
of Physics, University of Kashmir, Srinagar 190006, India}
\author{Subir Bhattacharyya}
\affiliation{Astrophysical Sciences Division, Bhabha Atomic Research Center, Mumbai 400085, India}
\affiliation{Homi Bhabha National Institute, Mumbai 400094, India}
\author{Debanjan Bose}
\affiliation{Department of Physics, Central University of Kashmir, Ganderbal 191201, India}
\begin{abstract}
The Flat Spectrum Radio Quasar (FSRQ) B2\,1308+326 was in its highest $\gamma$-ray flaring state during 60260-60310\,MJD. During this period, the source was detected in very high energy (VHE) by the large-sized telescope (LST-1). We conducted a detailed broadband spectral study of this source using the simultaneous data available in optical/UV, X-ray, and $\gamma$-ray bands. For the broadband spectral study, we select two gamma-ray high flux states (59750-59800\,MJD, 60260-60310\,MJD) and one low flux state (59250-59320\,MJD). During the epochs, 59750-59800\,MJD (high flux state) and 59250-59320\,MJD (low flux state), the broadband spectral energy distribution (SED) is well fitted using one zone leptonic emission model involving synchrotron, synchrotron self Compton (SSC) and external Compton (EC) processes. However, the flaring state (60260-60310\,MJD) during which the source showed VHE emission requires an additional component. We show that the inclusion of the photo-meson process can successfully explain this excess $\gamma$-ray emission. Further the estimated parameters, also suggest the source is transparent to VHE gamma-rays against pair production process.
\end{abstract}
\section{\label{sec:intro}Introduction}
Blazars are the class of active galaxies with strong evidence of a relativistic jet aligned towards the observer \citep{Antonucci-1993, urry-1995}.
The jet emission is predominantly non-thermal in nature and significantly Doppler-boosted \citep{Begelman-1984}.
Besides this non-thermal spectrum extending from radio--to--gamma-ray energies, certain blazars also exhibit broad emission/absorption line features \citep{Francis-1991, Liu-2006}, and accordingly, 
they are classified as FSRQs (with line features) and BL Lacs (with weak/no-emission lines) \citep{Padovani-2007}.
The broadband spectral energy distribution (SED) of blazars is characterized by two broad components \citep{Abdo-2010} with 
the low energy component, peaking at optical/UV--to--X-ray energies, and well understood to be synchrotron emission from a 
non-thermal electron distribution \citep{Blandford-1978, Maraschi-1992, Ghisellini-1993, Hovatta-2009}. 
The high-energy component, peaking at gamma-ray energies, is generally modeled as inverse Compton up-scattering of 
low-energy photons \citep{Ghisellini-1985, Begelman-1987, Blandford-1995}. However, the detection of neutrino from blazars and unusual gamma-ray spectral properties  
(e.g., orphan flares)
suggest the presence of hadronic emission in the high energy spectra of blazars. 

When the high energy emission is
attributed to the inverse Compton emission, the target photon field can be the synchrotron photons itself, commonly 
referred as synchrotron self Compton (SSC) emission \citep{Konigl-1981, Marscher-1985, Ghisellini-1989}, or the photon field external to the jet, referred as
external Compton (EC) \citep{Begelman-1987, Melia-1989, Dermer-1992}. The plausible external
photon field can be the broad emission lines (EC/BLR) \citep{Sikora-1994, Ghisellini-1996} or the IR photons (EC/IR) from the dusty torus proposed under 
unification theory \citep{Sikora-1994, Blajowski-2000, Ghisellini-2009}. Under these emission models, the hadrons are assumed to be cold and do not contribute to the 
radiative output. On the other hand, the models advocating the hadronic origin of the high energy emission involve proton synchrotron \citep{Aharonian-2000}
and cascades resulting from proton-proton and proton-photon interaction \citep{Mannheim-1989}. Besides these models, there also exist 
hybrid models where leptons and hadrons are both involved in the high-energy radiative processes and are often referred 
as lepto-hadronic models \citep{Diltz-2016, Paliya-2016}.
 
Blazars are further classified based on the peak frequency of their low-energy synchrotron spectral component: 
low-synchrotron-peaked (LSP) blazars with peak frequency $\nu_{\rm syn,peak}< 10^{14}$Hz; intermediate-synchrotron-peaked (ISP) 
blazars with $10^{14}<\nu_{\rm syn,peak}$ $< 10^{15}$Hz and high-synchrotron-peaked (HSP) blazars 
with $\nu_{\rm syn,peak}$ $> 10^{15}$Hz \citep{Abdo-2010}. FSRQs belong to the LSP blazar category. Some blazars also exhibit features of both BL Lacs and FSRQs during different flux states and are referred as transition blazars\citep{Ghisellini-2011, Ghisellini-2013, Ruan-2014}. 
Such blazars can be identified by investigating their broad-band SEDs (e.g., \citep{Ghisellini-2011,Ghisellini-2013}) and/or by determining the equivalent width (EW) of the broad emission lines in their optical spectra \citep{Ruan-2014}.

Blazar B2\,1308+326, also known as OP\,313, is one of the distant VHE blazar located at a redshift of 0.998 \citep{Hewett-2010}. 
Though its classification is uncertain \citep{Gabuzda-1993}, its nearly featureless optical spectrum \citep{wills-1979}, strong optical variability \citep{mufson-1985}
 and the significant optical polarisation \citep{Angel-1980} suggests this source is a BL Lac type \citep{Stickel-1991}. 
However, B2\,1308+326 also displays characteristics similar to those of a quasar. VLBI polarisation images reveal the polarised flux 
from the inner part of the jet is oriented perpendicular to the direction of the jet. This property is commonly associated with quasars 
rather than BL Lacs \citep{Gabuzda-1993}. The degree of core polarisation observed in B2\,1308+326 through VLBI measurements is relatively 
higher compared to typical quasars and falls towards the lower end of the spectrum for BL Lacs \citep{Gabuzda-1993}. B2\,1308+326 also 
exhibits an optical-UV excess beyond the extrapolation of the infrared \citep{brown-1989}. The large bolometric and the high luminosity at $5$ GHz \citep{Sambruna-1996, Kollgaard-1992} are further evidence that it could be treated as a quasar.

During December 2023, the source underwent a major $\gamma$-ray flare detected by Fermi and followed by the various optical facilities \citep{Bartolini-2023, Otero-Santos-2023}. During this period, LST-1 also witnessed an enhanced VHE emission with a significance greater than $5$ sigma \citep{Cortina-2023}. IceCube made a 30-day time window to search for the neutrino events originating from the direction of B2\,1308+326, but didn't find any significant neutrino detection from this source \citep{Thwaites-2022}.\\
Motivated by these observations, we used the simultaneous data available in Optical/UV, X-ray, and $\gamma$-ray bands to perform the broadband spectral study of a VHE-detected FSRQ B2\,1308+326 using the leptonic and hadronic processes. 
The manuscript is structured as follows: We discuss the data reduction procedure in \S \ref{observation}. The temporal study is described in \S \ref{temporal}. Details of SED modeling are provided in \S \ref{SED}. Cooling timescales of various radiative processes are discussed in \S \ref{cool_time}.  Results and our findings are summarized in \S \ref{summary}. A cosmology with $H_0$= 71 Km s$^{-1}$ Mpc$^{-1}$, $\Omega_m$ = 0.27, and $\Omega_{\Lambda}$ = 0.73 are used throughout this work.
\section{Observation and Data Reduction} \label{observation}
The FSRQ B2\,1308+326 has been detected at $\gamma$-ray energies by the \emph{Fermi} satellite since August 2008. Thanks to the wide-angle capabilities of \emph{Fermi}-LAT, the source was continuously monitored. However, \emph{Swift}, a pointing telescope, carried a total of 51 observations of the source up to January 2024 (60319\,MJD). For this study, we utilize the complete set of observations of the source made by \emph{Fermi} and \emph{Swift}, spanning from August 2008 (54682\,MJD) to Jan 2024 (60319\,MJD).
\subsection{\emph{Fermi}-LAT}
The \emph{Fermi} $\gamma$-ray telescope is a space observatory designed to observe the universe in a broad $\gamma$-ray energy spectrum. The Large Area Telescope (LAT), the main instrument of this system, observes photons with energies ranging from 20 MeV to 1 TeV through a pair production process. We utilized the $\gamma$-ray data of B2\,1308+326 obtained from the \emph{Fermi}-LAT during 54682-60319\,MJD. To make this data suitable for scientific analysis, we utilized the FERMITOOLS\footnote{\url{https://fermi.gsfc.nasa.gov/ssc/data/analysis/documentation/}}--v2.0.1 software for processing.
The standard data analysis procedure described in the \emph{Fermi}-LAT documentation\footnote{\url{https://fermi.gsfc.nasa.gov/ssc/data/analysis/}} is followed. We chose the Pass 8 Data (P8R3) within the 0.1-300 GeV energy range. We specifically focused on the SOURCE class events (evclass=128 and evtype=3) that fell within the $15^{\circ}$ region of interest (ROI) centered on the source position (RA: 197.619, Dec:32.3455). To prevent any interference from Earth limb $\gamma$-rays, the photons that come from zenith angle $>90^{\circ}$ are blocked. Furthermore, the latest version \emph{fermipy}-v1.0.1 \citep{Wood2017} is used for data reduction. In this work, we used the recommended model files for the Galactic diffuse emission component and extragalactic isotropic diffuse emission as gll$_-$iem$_-$v07.fits and iso$_-$P8R3$_-$SOURCE$_-$V3$_-$v1.txt respectively, and the post-launch instrument response function as P8R3$_-$SOURCE$_-$V3.
 
\subsection{\emph{Swift}-XRT/UVOT}
The \emph{Swift} satellite is equipped with three telescopes: the Burst Alert Telescope (BAT; 15-150 keV \citep{Barthelmy-2005}), the X-ray Telescope (XRT; 0.3-10 keV \citep{Burrows-2005}), and the UV/Optical Telescope (UVOT; 180-600 nm \citep{Roming-2005}). \\
During the period 54682-60319\,MJD, there are a total of 51 observations of B2\,1308+326 available in the X-ray and Optical/UV bands. We used XRTDAS V3.0.0, a software package included in the HEASOFT package (version 6.27.2), to process the X-ray data acquired in photon-counting (PC) mode. For XRT data, we obtained the cleaned event files using the standard XRTPIPELINE (version: 0.13.5). The XSELECT tool is used to choose source and background regions, and the corresponding spectrum files are obtained and saved, respectively. The source region is selected from a circular area of 20 pixels, while a circular area of 50 pixels away from the source location is chosen for the background. The XIMAGE is used to aggregate exposure maps, while the task xrtmkarf is employed to generate ancillary response files. Using the grppha task, the source spectra are grouped into bins to ensure that each bin contains a minimum of 20 counts. We used the XSPEC version 12.11.0 software \citep{Arnaud-1996} to fit the X-ray spectrum with a log-parabola/power-law model, considering absorption caused by neutral hydrogen (Tbabs). The neutral hydrogen column density $n_H$ value was fixed at $1.23\times 10^{20} cm^{-2}$ \citep{Kalberla-2005}; while the spectral indices and normalization of a log-parabola/power-law model were allowed to vary during spectral fitting.\\
The \emph{Swift}-UVOT telescope has also detected the FSRQ B2\,1308+326 in conjunction with the \emph{Swift}-XRT. This instrument is equipped with a total of 6 filters. Among them, three filters operate at optical wavelengths (V, B, and U), while the others operate at UV wavelengths (UVW1, UVM2, and UVW2). The UVOT data of B2\,1308+326 is processed into scientific products using the HEASOFT package (version 6.27.2). The UVOTSOURCE task included in the HEASOFT  was used to perform aperture photometry. We employed the uvotimsum task to combine multiple images in the filter. A circle with a radius of 5 arcsec has been selected to extract the source counts, while a circle with a radius of 10 arcsec has been used in a region near the target to estimate the background. The observed flux was corrected for the Galactic extinction using the $E(B-V)$ value of 0.0115 taken from \citet{Schlafly-2011}. The optical/UV spectra are fitted using XSPEC with an absorbed power-law model.

\section{Temporal Analysis} \label{temporal}
\begin{figure*}
\centering
\includegraphics[scale=0.4]{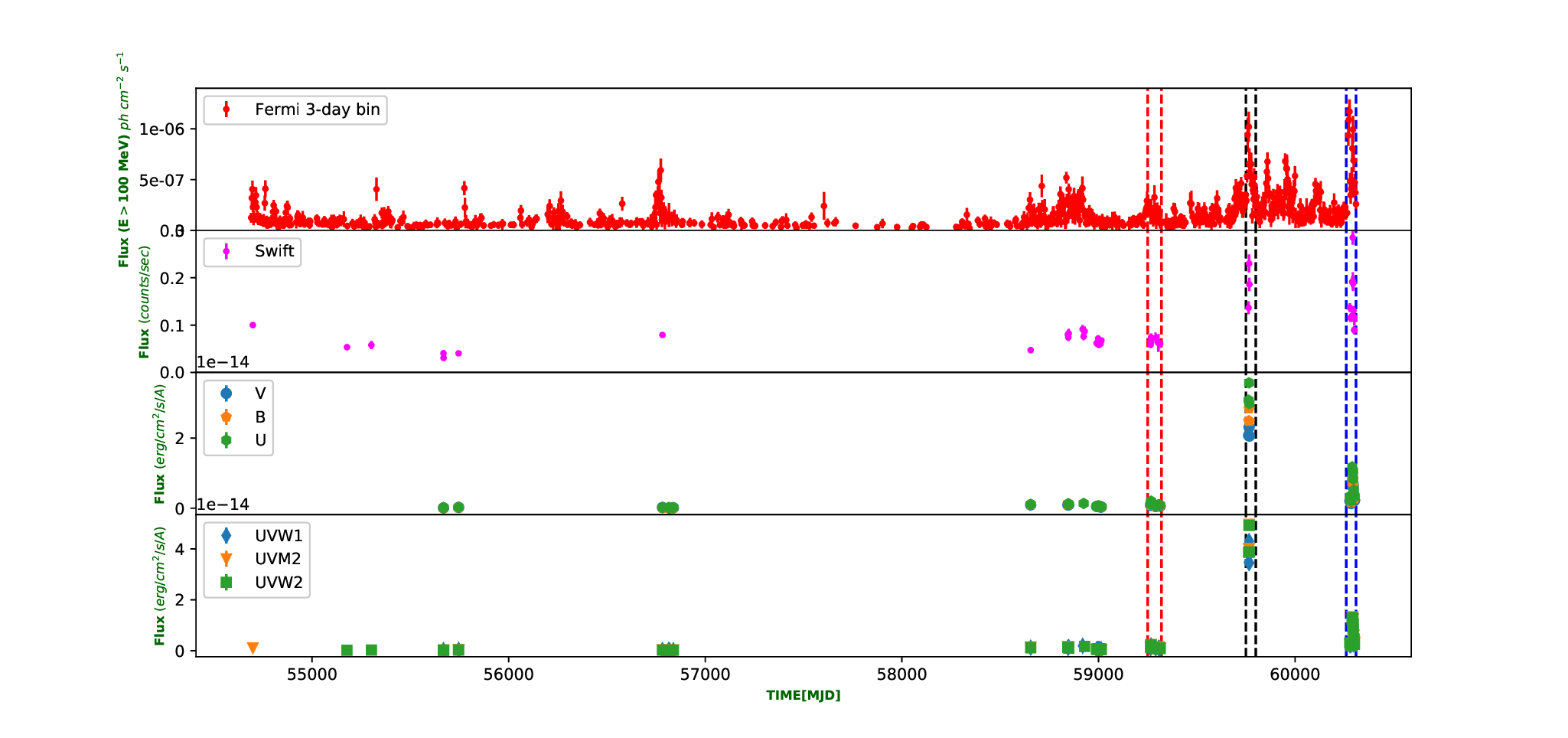}
\caption{Multi-wavelength lightcurve of B2\,1308+326 in different flux states. The top panel of the multiplot displays the 3-day binned $\gamma$-ray lightcurve integrated over the energy range of 0.1-100 GeV, the upper middle panel displays the X-ray lightcurve in the energy range of 0.3-10 keV, the lower middle panel and bottom panel displays Optical and UV lightcurves respectively.}
 \label{fig:MLC}
\end{figure*}

To study the temporal behavior of B2\,1308+326, we acquired the 3-day binned $\gamma-$ray lightcurve from the \emph{Fermi}-LAT Light Curve Repository \citep{Abdollahi-2023}, covering the time range 54682-60319\,MJD. Figure $\ref{fig:MLC}$ displays the multiwavelength light curve of B2\,1308+326, obtained through observations from \emph{Fermi}-LAT and \emph{Swift}-XRT/UVOT. The $\gamma$-ray light curve shows various flaring activity, with simultaneous activity in optical/UV and X-ray energy bands for some epochs. For the spectral study, we chose two flaring and one quiescent state, which are shown by dashed vertical lines in Figure $\ref{fig:MLC}$, and the details of these states are given in Table $\ref{tab:start_end_dates}$. 

\begin{table*}[!htb]
\caption{Details of the selected flux states: Col:- 1: States, 2: Start date, 3: End date, 4: Nature of the states.}
     \centering
     \vspace{0.2cm}
\begin{tabular}{ l c c c}
\hline
\hline
State& Start Date (MJD)& Stop Date (MJD) &  Activity \\
\hline 
State I & 2021-Feb-05 (59250)& 2021-April-16 (59320)&  Low Flux \\
State II & 2022-June-20 (59750)& 2022-Aug-09 (59800)& High Flux\\
State III & 2023-Nov-12 (60260)& 2024-Jan-01 (60310)& High Flux\\
\hline
\hline
\label{tab:start_end_dates}
\end{tabular}
\end{table*}
The source shows maximum $\gamma$-ray and X-ray activity during  60260-60310\,MJD, with integrated fluxes $\rm 1.17\times 10^{-6}\,ph\,cm^{-2} s^{-1}$ and $\rm 2.85 \times 10^{-1}\,counts\,s^{-1}$ respectively. However, there was no flaring activity in optical/UV during this epoch. Furthermore, the source showed high activity in $\gamma$-ray, X-ray, and optical/UV energy bands during 59750-59800\,MJD. These correlated/uncorrelated flux enhancements observed from the multiwavelength light curve
suggest different emission processes active in these energy bands. To measure the extent of variation in the source over time, we compute the fractional variability amplitude of the source in different energy bands using \citep{Vaughan-2003}. 
\begin{equation}
    \label{eq3}
   \rm F_{var}=\sqrt{\frac{S^2-\overline{\sigma_{err}^2}}{\overline{F}^2}}
\end{equation}
Here, $\rm S^2$ is the variance, $\rm \overline{F}$ is the mean flux, and $\rm \overline{\sigma_{err}^2}$ the mean square of the measurement error on the flux points. The uncertainty on $\rm F_{var}$ is given by \citet{Vaughan-2003}
\begin{equation}
    \label{eq4}
    \rm F_{var,err}=\sqrt{\frac{1}{2N}\left(\frac{\overline{\sigma_{err}^2}}{F_{var}\overline{F}^2}\right)^2+\frac{1}{N}\frac{\overline{\sigma_{err}^2}}{\overline{F}^2}}
\end{equation}
N is the number of flux points in the light curve across all energy bands. The results of this variability analysis are shown in Table $\ref{tab:Fractional_variability}$. 
The lowest value of $F_{\rm var}$ is witnessed for X-ray energies than the other two energy bands. On the other hand, the source showed higher fractional variability amplitudes in the low energy band (optical/UV) compared to the high energies (X-ray/$\gamma$-ray). These results are consistent with the previous studies \citep{Pandey-2024}.
\begin{table}[!htb]
\caption{Fractional Variability amplitude ($\rm F_{\rm var}$) of source in different energy bands with simultaneous data across the light curve. }
    \centering
    \begin{tabular}{lc}
    \hline
     \hline
     Energy band    & $\rm F_{\rm var}$ \\
     \hline
       $\gamma$-ray (0.1 - 100 GeV)  & $0.831 \pm 0.013$ \\
       X-ray (0.3 - 10 keV)  & $0.559 \pm 0.014$ \\
       UVW2   & $1.759 \pm 0.005$  \\
       UVM2  & $2.204 \pm 0.006$ \\
       UVW1  & $1.629 \pm 0.006$ \\
       U     & $1.593 \pm 0.005$  \\
       B   &  $1.585 \pm 0.005$\\
       V    & $1.779 \pm 0.006$  \\
       \hline
        \hline
       \label{tab:Fractional_variability}
    \end{tabular}
\end{table}

\section{Broadband Spectral Modeling} \label{SED}
To understand the spectral behavior of the source during various flux states, we model the optical/UV, X-ray, and $\gamma$-ray spectra of the source using emprical functions.
The $\gamma$-ray data in different flux states are modeled with power-law (PL) and log-parabola (LP) models. The PL model is defined as 
\begin{equation}
    \label{eq5}
    \frac{dN}{dE} = N_0 \left(\frac{E}{E_0}\right) ^{-\Gamma}
\end{equation}
where, $\rm N_0$ represents the normalisation and $\Gamma$ denotes the slope or PL spectral index.\\
The LP model is defined as 
\begin{equation}
    \label{eq6}
        \frac{dN}{dE} = N_0 \left(\frac{E}{E_0}\right) ^{-(\alpha + \beta \log(E/E_0))}
\end{equation}
where, $\rm N_0$ is the normalization corresponding to the differential number density at $\rm E_0$, $\alpha$ is the spectral slope at $\rm E_0$ and $\beta$ is spectral curvature parameter.
The best-fit parameters are listed in the Table $\ref{tab:gamma_spec}$.
The spectral points used for broadband SED modeling are obtained from the log-parabola fit when there is a significant curvature in the $\gamma$-ray spectrum. To determine the significance of  curvature in the considered flux states, we computed the curvature test statistic $TS_{\rm curve}=2[\log\mathcal{L}(LP)-\log\mathcal{L}(PL)$] \citep{Nolan-2012}. Curvature is considered significant if $TS_{\rm curve}>16$. $TS_{\rm curve}$ suggests that there is a significant curvature in State II ($TS_{\rm curve}=28$) and State III ($TS_{\rm curve}=62.52$). However, the $TS_{\rm curve}$  value for State I is obtained as 8.28, indicating no significant curvature. Based on $TS_{\rm curve}$ value, we choose the $\gamma$-ray SED points for two flaring states from the log-parabola fit and for the low-flux state from the power-law fit.
\begin{table*}
\caption{Results of the model fit to the $\gamma$-ray spectra of B2\,1308$+$326, obtained for different time periods. Col.[1]: period of observation (MJD); Col.[2]: activity state; Col.[3]: the model used (LP: logParabola); Col.[4]: integrated $\gamma$-ray flux (0.1$-$300 GeV), in units of 10$^{-6}$ ph cm$^{-2}$ $s^{-1}$, Col.[5]: Spectral index; [6]: Curvature index; [7]: Test statistics (TS); [8]: -log(likelihood); [9]: Curvature parameter.}
\label{tab:gamma_spec}
\begin{center}
\begin{tabular}{ccccccccc}
\hline\hline
Period & Activity & Model & $F_{0.1-300~{\rm GeV}}$ & $\alpha$/$\Gamma$ & $\beta$ & TS & -$\log\mathcal{L}$ & $TS_{\rm curve}$ \\
~[1] & [2] & [3] & [4] & [5] & [6] & [7] & [8] & [9]\\
\hline
59250$-$59320 & Low Flux & LP & 0.08 $\pm$ 0.03 & 2.32 $\pm$ 0.15 &  0.20 $\pm$  0.09 & 203.90 & 13615.30 & 8.28              \\
              &          & PL & 0.12 $\pm$ 0.02 & 2.17 $\pm$ 0.08& -- & 230.84 & 13619.44 & --\\
            
59750$-$59800 & High Flux      & LP & 0.49 $\pm$ 0.05 & 2.14 $\pm$ 0.04 &  0.09 $\pm$  0.02  & 3206.30 & 15585.40 & 28            \\
              &          & PL & 0.56 $\pm$ 0.02 & 2.04 $\pm$ 0.02 & -- & 4094.16 & 15599.40 & --\\
               
60260$-$60310 & High Flux      & LP & 0.45 $\pm$ 0.05 & 1.74 $\pm$ 0.04 &  0.11 $\pm$ 0.02 & 4069.12 & 18474.82 & 62.52           \\
&          & PL & 0.71 $\pm$ 0.02 & 1.83 $\pm$ 0.02 & -- & 9760.87 & 18506.08 & --\\
              
\hline
\end{tabular}
\end{center}
\end{table*}

Fifty one observations were available in \emph{Swift} during the time interval 54682-60319\,MJD. To generate the X-ray spectrum, we used the ``xselect" tool to obtain the source and background files of each observation ID. An ancillary response function (ARF) has been generated using the tool ``xrtmkarf". The ``grppha" task was utilized to obtain 20 counts per bin. The spectra are then fitted using the X-ray spectral-fitting package ``XSPEC" with PL and LP.
The X-ray spectra of two flaring states are well-fitted using the tbabs*LP model with $\chi^2$/dof 18.79/17 (State II) and 97.87/99 (State III), whereas the tbabs*PL model yields a better fit for the spectra of the low flux state with $\chi^2$/dof 13.02/18 (State I). For UVOT, we used the ``uvotimsum" tool to merge the images from the individual filters in the selected flux state. Then, we obtained the flux values for each filter using the ``uvotproduct" task. 

\begin{figure}
   \centering
   \includegraphics[width=0.35\textwidth,angle=270]{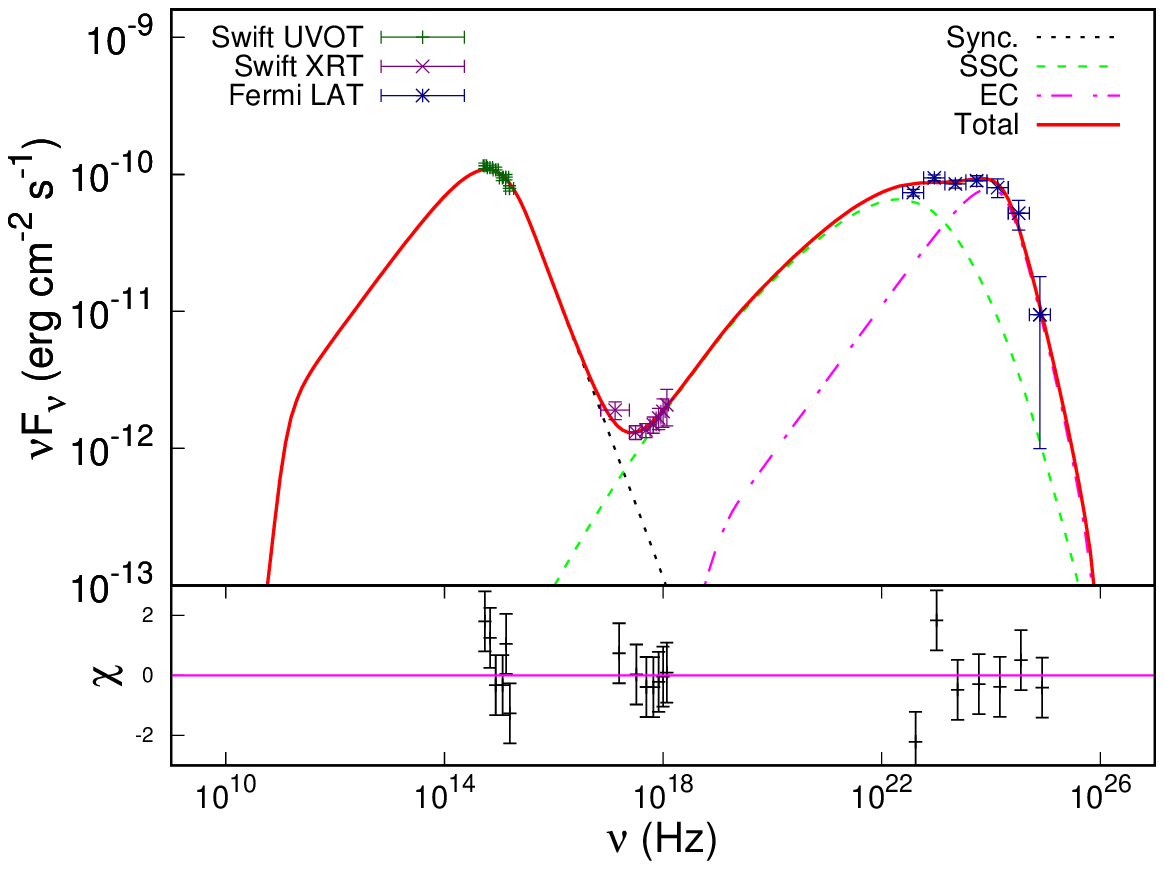}
   \vspace{0.5cm}
   \caption{ Spectral fit to the broad-band SED of B2\,1308+326 during the epoch 59750-59800\,MJD. The left and right keys illustrate the 
   data points and emission mechanisms.
   \label{fig:SED_E1}}
\end{figure}

\begin{figure}
   \centering
   \includegraphics[width=0.35\textwidth,angle=270]{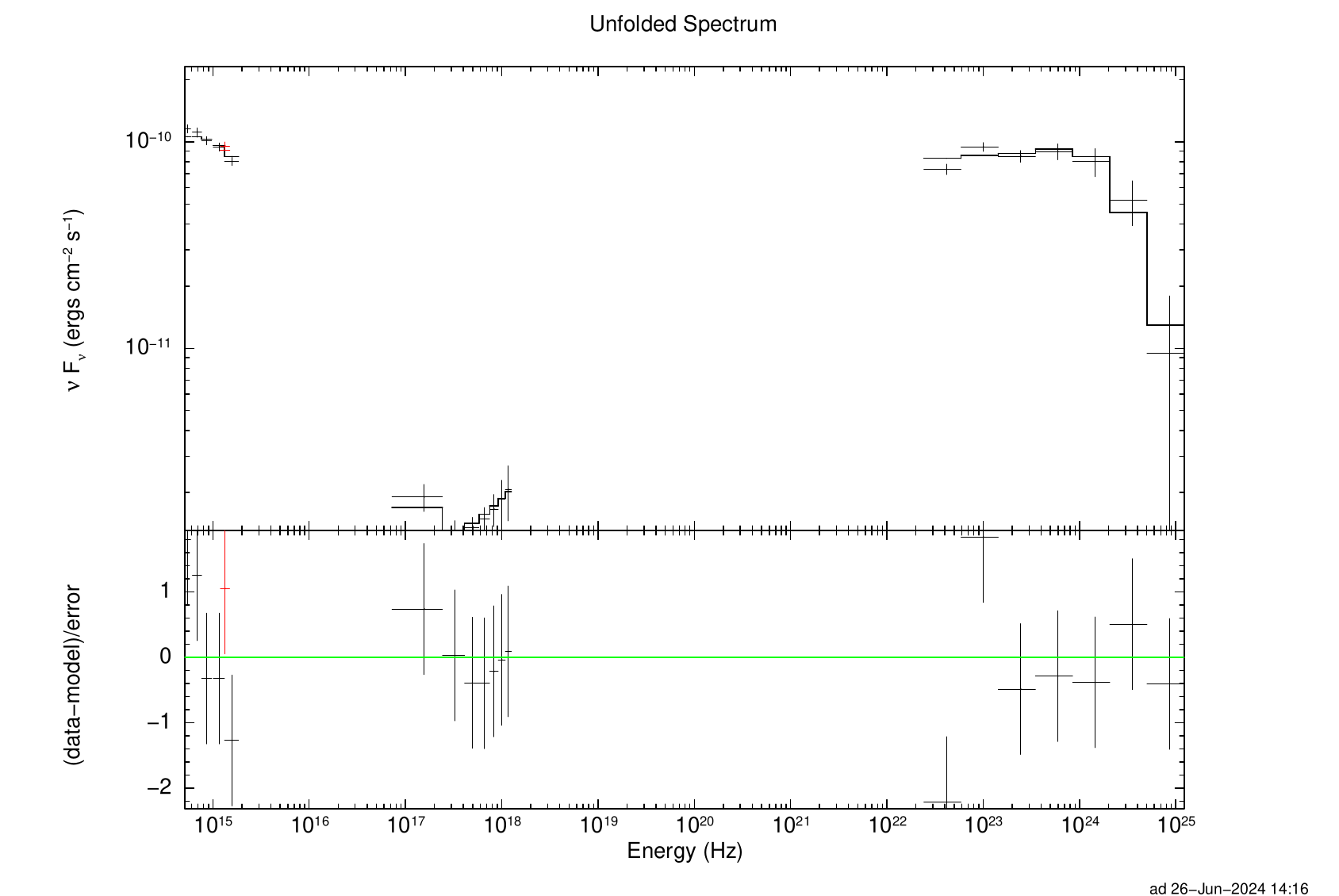}
   \vspace{0.5cm}
   \caption{Fitting data of B2\,1308+326 with the one zone leptonic emission model during the epoch 59750-59800\,MJD using XSPEC.
   \label{fig:xspec_F1}}
\end{figure}

\begin{figure}
   \centering
   \includegraphics[width=0.35\textwidth,angle=270]{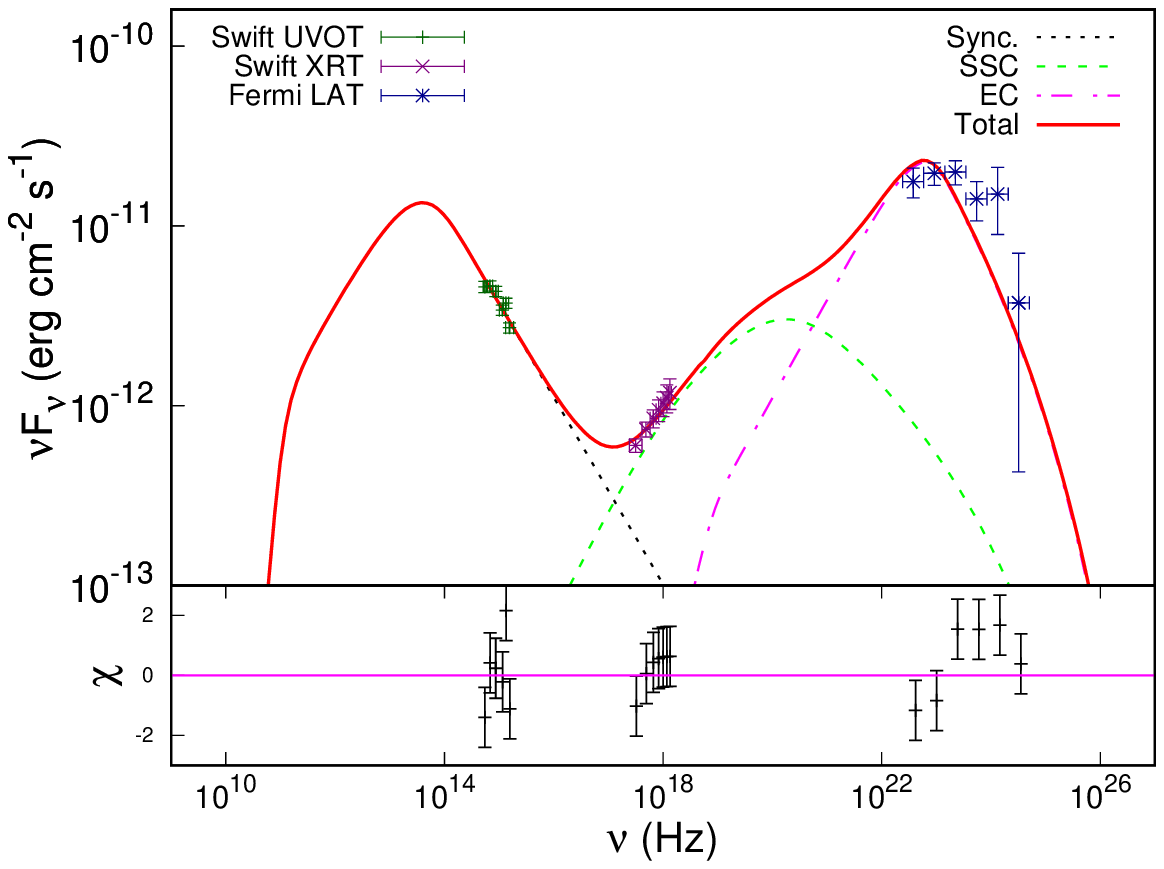}
   \vspace{0.5cm}
   \caption{Spectral fit to the broad-band SED of B2\,1308+326 during the epoch 59250-59320\,MJD. The left and right keys illustrate the 
   data points and emission mechanisms.
   \label{fig:SED_EQ}}
\end{figure}

\begin{figure}
   \centering
   \includegraphics[width=0.35\textwidth,angle=270]{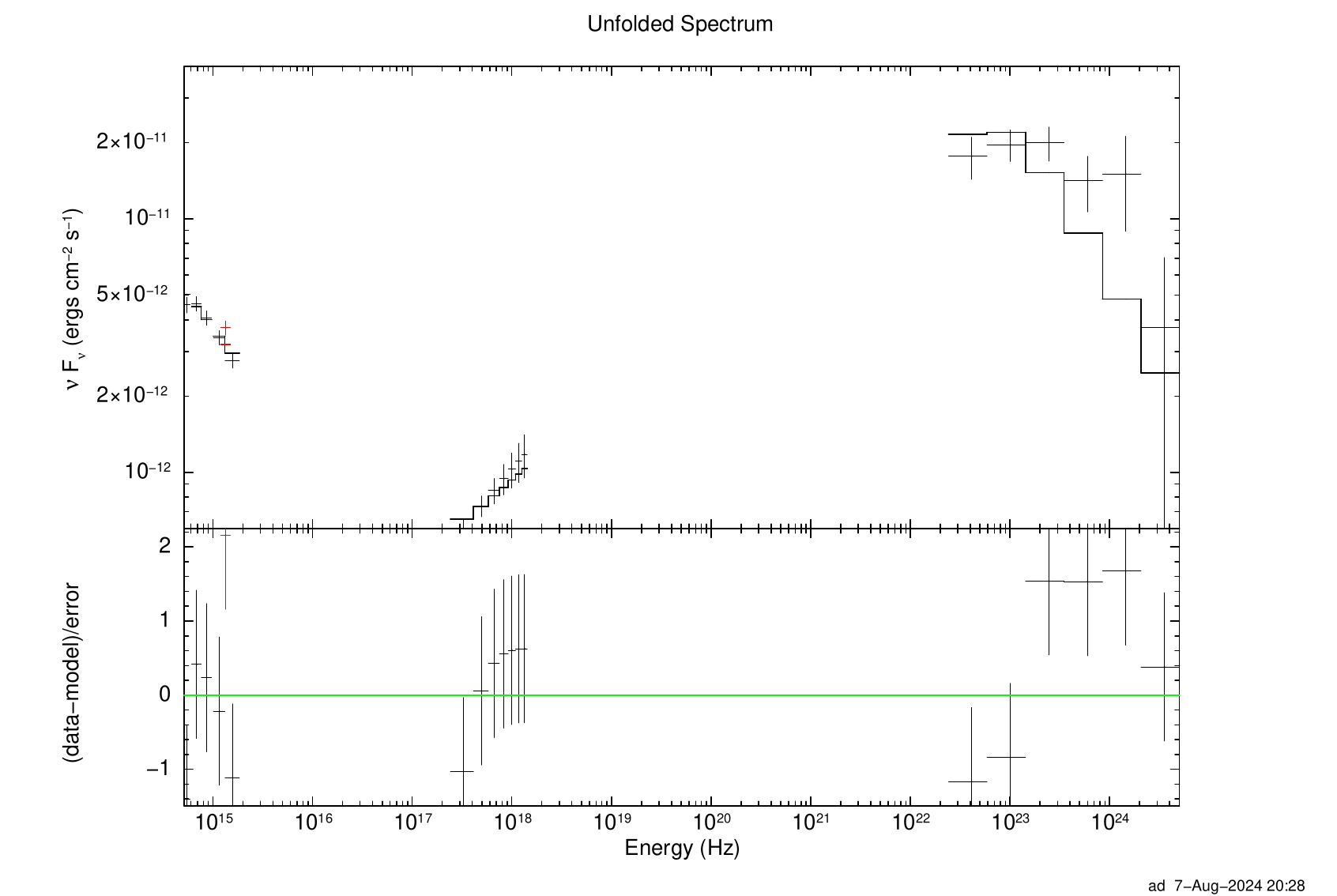}
   \vspace{0.5cm}
   \caption{Fitting data of B2\,1308+326 with the one zone leptonic emission model during the epoch 59250-59320\,MJD using XSPEC.
   \label{fig:xspec_Q}}
\end{figure}

\begin{figure}
   \centering
   \includegraphics[width=0.35\textwidth,angle=270]{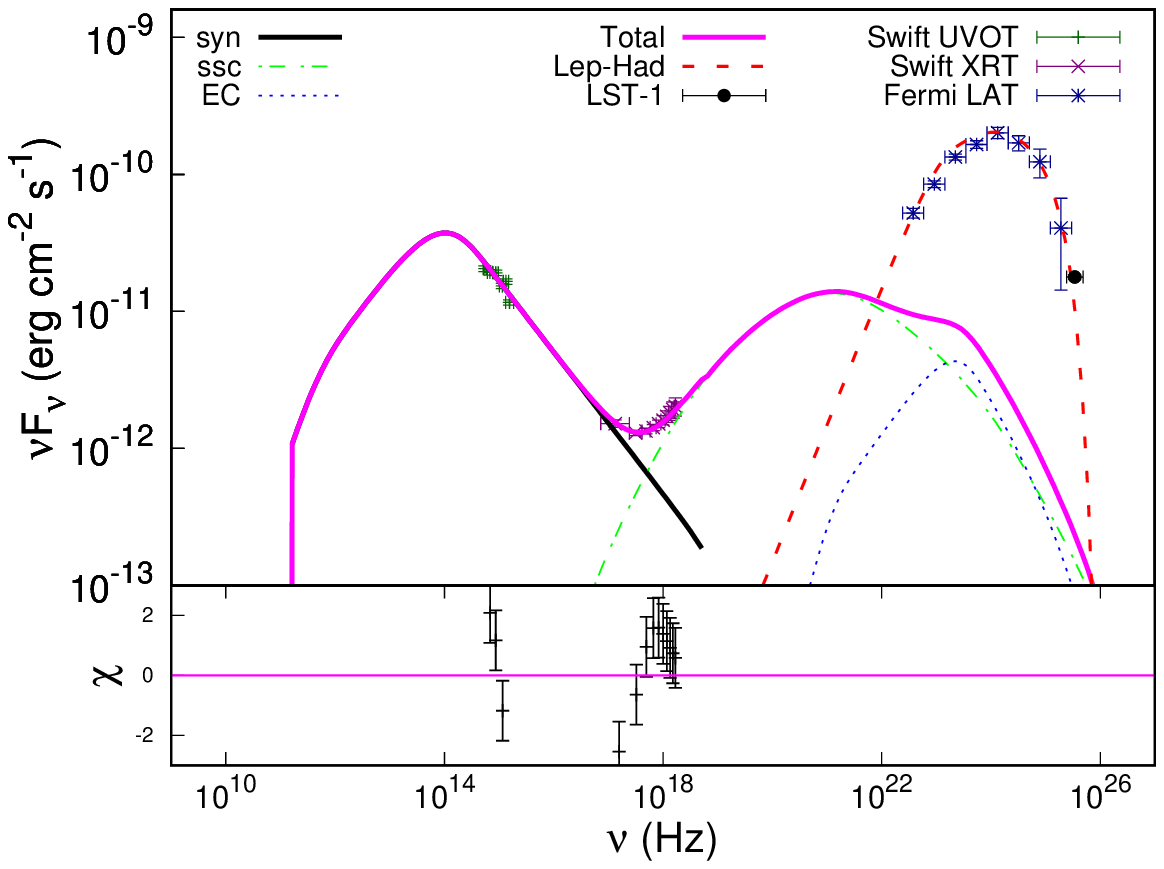}
   \vspace{0.5cm}
   \caption{Broad band SED of B2\,1308+326 with the one zone leptonic emission model fits along with the photo-meson process during the epoch
    60260-60310\,MJD with LST-1 flux point included. The right and left keys illustrate the emission mechanisms and data points.
   \label{fig:SED_LH}}
\end{figure}

\begin{figure}
   \centering
   \includegraphics[width=0.35\textwidth,angle=270]{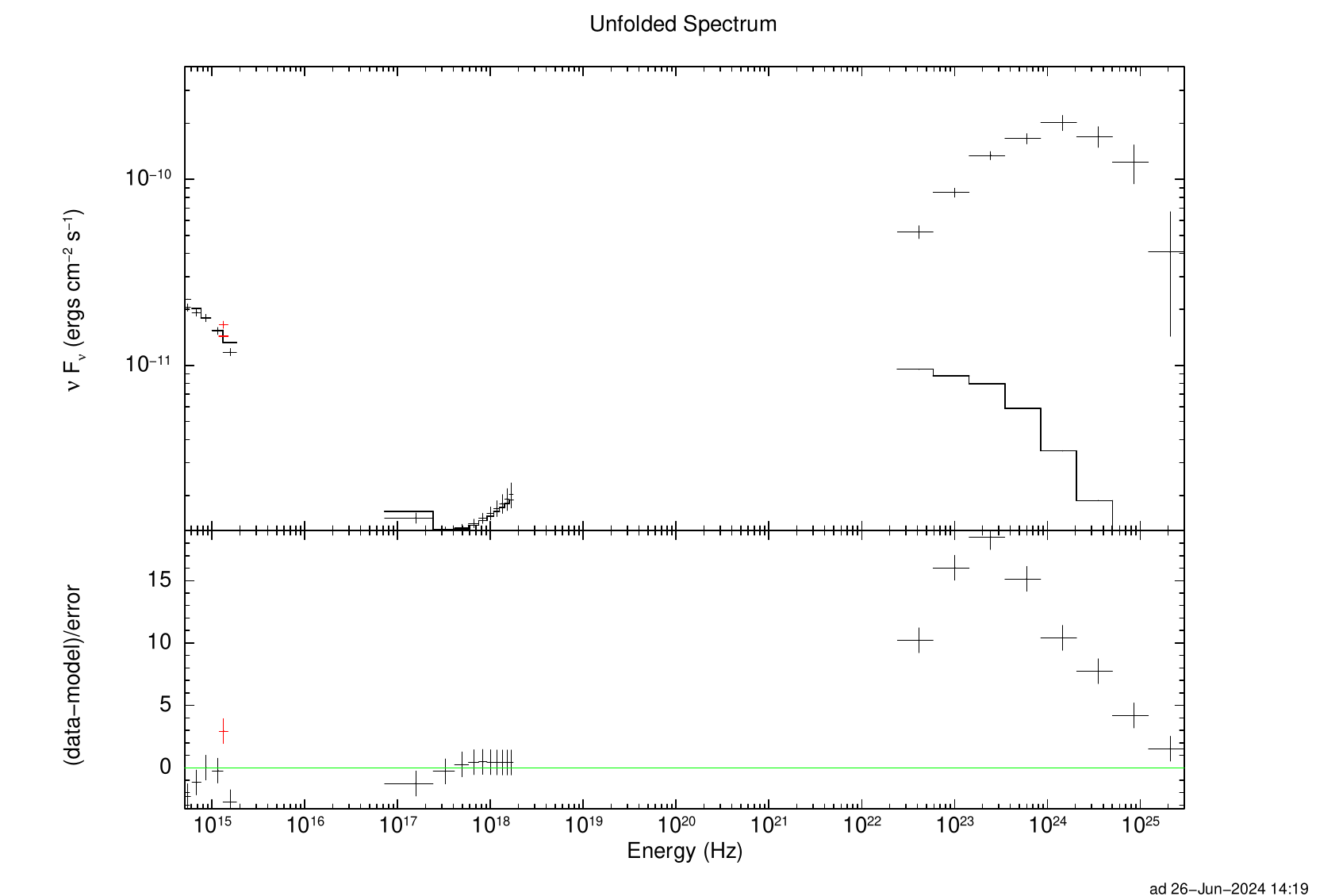}
   \vspace{0.5cm}
   \caption{Fitting data of B2\,1308+326 with the one zone leptonic emission model during the epoch 60260-60310\,MJD using XSPEC.
   \label{fig:xspec_F2}}
\end{figure}

\begin{figure}
   \centering
   \includegraphics[width=0.35\textwidth,angle=270]{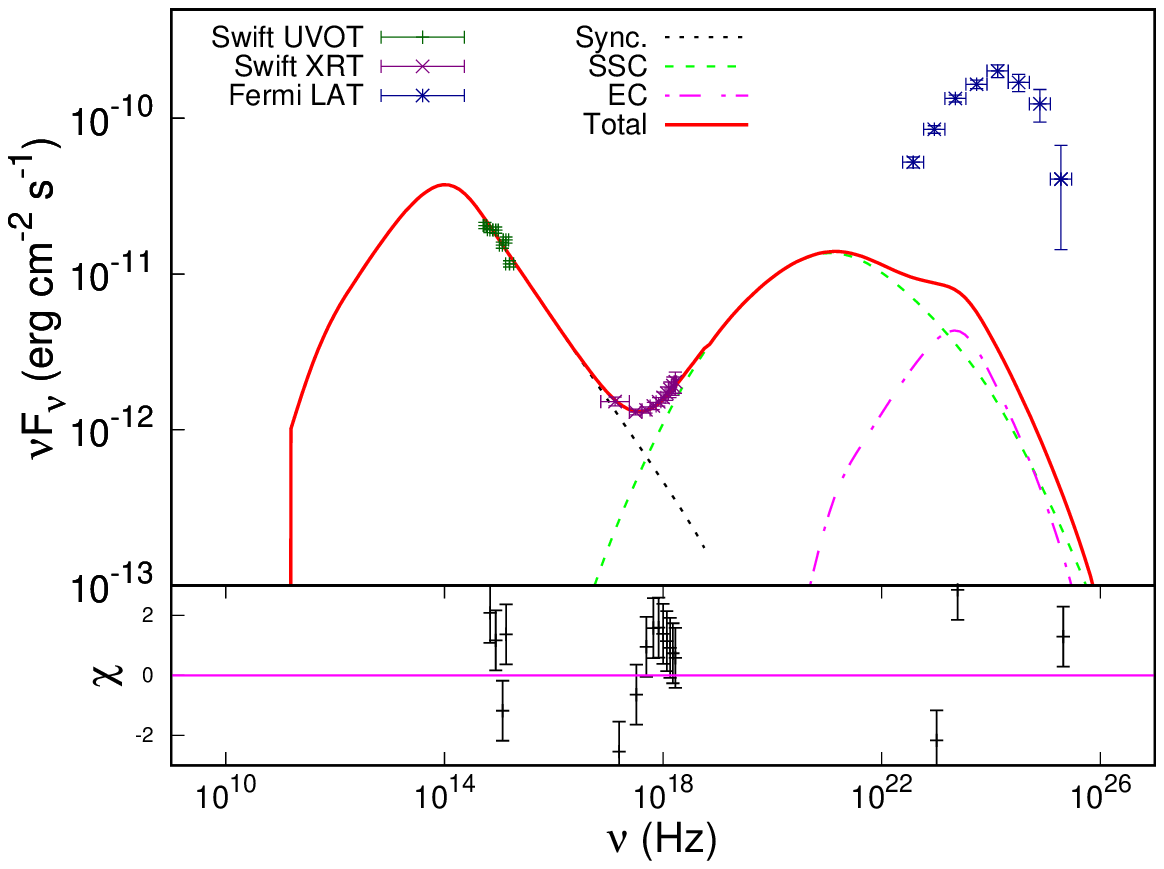}
   \vspace{0.5cm}
   \caption{Broad band SED of B2\,1308+326 with the one zone leptonic emission model during the epoch 60260-60310\,MJD.
   \label{fig:VHE_LEP}}
\end{figure}

The variations in the spectral shape can provide hints regarding the underlying emission mechanism. To facilitate this, 
we performed a broadband spectral fitting of B2\,1308+326
during different flux states using synchrotron, SSC, and EC emission mechanisms. We chose three epochs with simultaneous multiwavelength data available, and among them,the source was in a high flux state during two epochs, 59750-59800\,MJD and 60260-60310\,MJD, while in a low flux state during  59250-59320\,MJD. The broadband SED is modeled using a one-zone leptonic emission model \citep{Sahayanathan-2012, shah_2017, Sahayanathan-2018} where the emission region is assumed to be a spherical blob with a radius $R$ and populated with a broken power-law electron distribution given by
\begin{equation}
 N(\gamma)\;d\gamma = K \times
\begin{cases}
\gamma^{-p}d\gamma& {\rm for  } \quad\gamma_{\rm min} <\gamma<\gamma_{\rm b}\\
\gamma^{ q-p}_{\rm b}\gamma^{-q}d\gamma &{\rm for} \quad \gamma_{\rm b} < \gamma < \gamma_{\rm max}
\end{cases} \quad cm^{-3}
\label{equation:bkn}
\end{equation}
where, $K$ is the normalization factor, $\gamma$ is the electron Lorentz factor with $\gamma_{\rm min}$ and $\gamma_{\rm max}$ 
denoting its minimum and maximum value, $\gamma_{\rm b}$ is the Lorentz factor corresponding to the break in the electron distribution, and $p$ and $q$ 
are the low and high energy indices of the particle distribution. We assume the emission region is permeated with the tangled magnetic field $B$ and moves down the jet with a bulk Lorentz factor $\Gamma$ at an angle $\theta$ relative to the observer. The electron distribution described in equation $\ref {equation:bkn}$ will dissipate energy through synchrotron, SSC, and EC emission mechanisms. Based on the location of the emission region from the central engine, the primary external photon field can either be Ly-$\alpha$ line emission from the BLR (EC/BLR) or thermal IR photons (EC/IR) from the dusty torus \citep{Ghisellini-2009}. The emissivities resulting from these radiative processes are calculated numerically, and the flux observed on earth is determined by considering relativistic and cosmological effects \citep{Begelman-1980, Dermer-1995}. 
The numerical code involving synchrotron, SSC, and EC emission process was added as a local model in XSPEC and used to fit the broadband SED for the chosen epochs (59750-59800\,MJD, 60260-60310\,MJD, and 59250-59320\,MJD). The $n_{H}$-corrected source X-ray flux is calculated using the cpflux tool in the XSPEC. The optical/UV flux may have contamination and may deviate largely from a power law. 
To account for this, we introduced additional systematic errors to the optical data so that a power-law spectral fit can result in reduced $\chi^2 \sim$ 1 \citep{Dar-2024}.
This additional systematic error was chosen to be $4$ \% for two flaring states and $6$ \% for the quiescent state. The ASCII data, which includes corrected X-ray, optical/UV, and $\gamma$-ray fluxes, were then converted into a Pulse Height Analyser (PHA) file using the HEASARC (High Energy Astrophysics Science Archive Research Center) tool flx2xsp. 
The observed broadband spectrum is primarily governed by $10$ parameters: $K$, $p$, $q$ and $\gamma_{\rm b}$ describing the electron distribution, $\Gamma$, $B$, $R$ and $\theta$ describing the macroscopic emission region properties and, the rest two parameters describing the external photon field frequency ($\nu_*$) and energy density $U_*$. 
For numerical simplicity, the external photon field is assumed to be a black body at temperature $T$ and its energy density $U_* = f\; U_{bb}$ where, $U_{bb}$ is the black body energy density, and $f$ is the fraction involved in the inverse Compton scattering process. 
Also, we assume equipartition between the magnetic field and the electron energy densities \citep{Burbidge-1995, Kembhavi-1995}.  
Initial fit was performed with these 10 parameters; however, due to possible degeneracies, the confidence level was obtained only for the four parameters: $p$, $q$, $\Gamma$, and $B$. 
Our spectral fit suggests during 59250-59320\,MJD and 59750-59800\,MJD epochs, the $\gamma$-ray spectrum can be better explained by the EC scattering of IR photons emanating from dust at temperature 1000 K. The best-fit parameters are listed in Table $\ref{tab:best-fit_parameters}$, and the best-fit model with the observed data and residue are shown in Figures $\ref{fig:SED_E1}-\ref{fig:xspec_F2}$. Interestingly, during the brightest $\gamma$-ray flare (60260-60310\,MJD), when the source was also detected in VHE, the $\gamma$-ray spectrum cannot be explained satisfactorily by the EC model as shown in Figure $\ref{fig:VHE_LEP}$. In this figure, we have fixed $\gamma_b$ at 2500, but even when considering a range of $\gamma_b$ values, the EC model still fails to satisfactorily reproduce the $\gamma$-ray spectrum.
To investigate this further, we included an additional emission component resulting from the photo-meson process. Particularly, we consider the interaction of relativistic protons with the synchrotron photons, where the later is assumed to be a power-law. 
We followed the procedure described in \citet{Kelner-2008} to obtain the $\gamma$-ray spectrum resulting from the decay of  $\pi^{0}$-meson.  The differential $\gamma$-ray photon number density can be obtained from
\begin{equation}
	\frac{dN_\gamma}{dE_\gamma} =  \int f_\text{p} (E_\text{p}) f_{ph}(\epsilon) \phi_\gamma(\eta, \text{x}) \frac{dE_{\text{p}}}{E_{\text{p}}} d\epsilon
\end{equation}
where, $f_p(E_p) \, dE_p$ and $f_{\text{ph}}(\epsilon) \,d\epsilon$ are the number densities of protons and photons, 
$\Phi_{\gamma}(\eta, x)$ is the angle averaged cross-section of the p$\gamma$-interaction. $\eta$ and $x$ are the dimensionless quantities given by \citet{Kelner-2008}
\begin{equation}
\eta = \frac{4\,\epsilon E_p}{m_p^2 c^4}\,, \qquad x = \frac{E_\gamma}{E_p}
\end{equation}

We found that with the addition of this hadronic emission component, the $\gamma$-ray spectrum can be reproduced well.
Through this broadband spectral modeling of B2\,1308+326, we conclude that the blazar jet may involve a significant hadronic emission mechanism; however, it may be dominant at $\gamma$-ray energies during a major flare. The optical spectrum can be well explained by the synchrotron emission from the leptons during all flux states. The X-ray spectrum, on the other hand, is largely due to inverse Compton emission during the low flux states. During high flux states, the X-ray spectrum falls in the transition regime between the synchrotron and inverse Compton spectral components. This indicates that the spectrum shifts towards the bluer end during high flux states. Such spectral behavior can be associated with the increase in the bulk Lorentz factor of the jet. Interestingly, our broadband spectral fit also suggests that the high flux states are associated with large jet Lorentz factors. Nevertheless, we cannot assert this inference based on the spectral modeling at three specific flux states, which also involves a significant number of parameters that cannot be constrained.

\section{Radiative cooling timescales}\label{cool_time}
To further investigate the emission mechanisms during the flaring state corresponding to 60260-60310 MJD, we calculated the cooling timescales of various radiative processes \citep{Rybicki-1979}. Understanding these timescales would put additional constraints on the processes responsible for the observed broadband spectrum during the flaring period. We first calculated the synchrotron cooling timescale which is given as $ t_{\text{syn}} = \frac{6\pi m_e c}{\sigma_T B^2 \gamma_e}$ \citep{Rybicki-1979}, where $m_e$ is the electron mass, $c$ is the speed of light, $\sigma_T$ is the Thomson scattering cross-section, $B$ is the magnetic field strength, $\gamma_e$ is the Lorentz factor of the electron. Since the synchrotron cooling time is inversely proportional to electron energy and magnetic field strength, high-energy electrons in a strong magnetic field experience rapid cooling, which is crucial for shaping the observed emission at optical/UV and soft X-ray bands. The SSC cooling time is given by $t_{\text{SSC}} = \frac{3m_e c}{4 \sigma_T \gamma_e ({U_{\text{B}} + \kappa_{\rm KN}U_{syn}})}$ \citep{Rybicki-1979}, where $U_{\text{syn}}$ is the energy density of the synchrotron radiation field, $U_{\text{B}} = \frac{B^2}{8\pi}$ is the energy density of the magnetic field. SSC cooling becomes significant in environments with low magnetic fields, and $\kappa_{\rm KN}$ is the energy-dependent correction factor to cross-section accounting for Klein-Nishina effects given as 
\begin{equation}
\begin{split}
\kappa_{\rm KN} = \frac{3}{4} \bigg[ \frac{1+x}{x^3} 
 \bigg( \frac{2x(1+x)}{1+2x} - \ln(1+2x) \bigg) \\
  + \frac{1}{2x} \ln(1+2x) - \frac{(1+3x)}{(1+2x)^2} 
  \bigg]
\end{split}
\end{equation}
Where $x \equiv \frac{h\nu}{mc^2}$
 \citep{Rybicki-1979}. 
The EC energy loss timescale is given by: $t_{\text{EC}} = \frac{3m_e c}{4 \sigma_{T} \kappa_{\rm KN} \gamma_e U_{\rm ph}}$. Where $U_{\rm ph}$ is the target photon field (i.e, Thermal IR photon field).
The EC cooling time mainly depends on the external photon field and is typically significant in the presence of strong external photon sources such as the broad-line region or dusty torus.\\ The hadronic cooling timescale, such as proton-synchrotron cooling, is given by $t_{\text{p, syn}} = \frac{6 \pi m_e c}{\sigma_T \gamma_p B^2} \left( \frac{m_p}{m_e} \right)^3$ where, $m_p$ is the proton mass and $\gamma_p$ is the proton energy. 
The p$\gamma$ energy loss timescale is obtained using \citep{Dermer-2009}
\begin{equation}
 t_{\rm p \gamma}(\gamma_p) =
\left[\int_{\frac{\epsilon_{\rm th}} {2 \gamma_p}}^{\infty}{\rm d}\epsilon\;{
\frac{c \, n_{\rm ph}(\epsilon)}{2 \gamma_p^2 \epsilon^{ 2} }
\int_{\epsilon_{\rm th}}^{2\epsilon \gamma_p}{\rm d} \epsilon_{\rm r}\;{
\sigma(\epsilon_{\rm r}) K_{\rm p \gamma}(\epsilon_{\rm r}) \epsilon_{\rm r} 
  }\, }\right]^{-1}
\label{tpgiso}
\end{equation}
Here, $\gamma_{p}= \frac{E_{p}}{m_{p}c^2}$ is the proton Lorentz factor, $\epsilon=\frac{h\nu}{m_{e}c^2}$, and $\epsilon_{r}$ is the photon energy in the rest frame of proton.
For simplicity, we approximate the cross-section
$\sigma(\epsilon_{\rm r})$ as a sum of 2 step-functions
$\sigma_{1}(\epsilon_{\rm r})$ and $\sigma_{2}(\epsilon_{\rm r})$ corresponding to the single-pion and multi-pion production channels, respectively. For the single-pion channel, $\sigma_{1}=340~ \mu \rm b$ for $ \epsilon_{th} = 390 \leq
\epsilon_{\rm r}
\leq 980$ and $\sigma_1 = 0$ outside this region, whereas, for the multi-pion channel, $ \sigma_2 = 120 ~\mu \rm b$ at $\epsilon_{\rm r} \geq 980 $. The inelasticity is approximated as $K_{\rm
p\gamma} =K_1\approx 0.2$ for the single-pion channel ($ 390 \leq \epsilon_{\rm r} \leq 980$) and $K_{p\gamma}=K_2\approx 0.6$ for energies above 980 \citep{Dermer-2009}. 

\begin{figure*}
   \centering
   \includegraphics[width=1.0\textwidth]{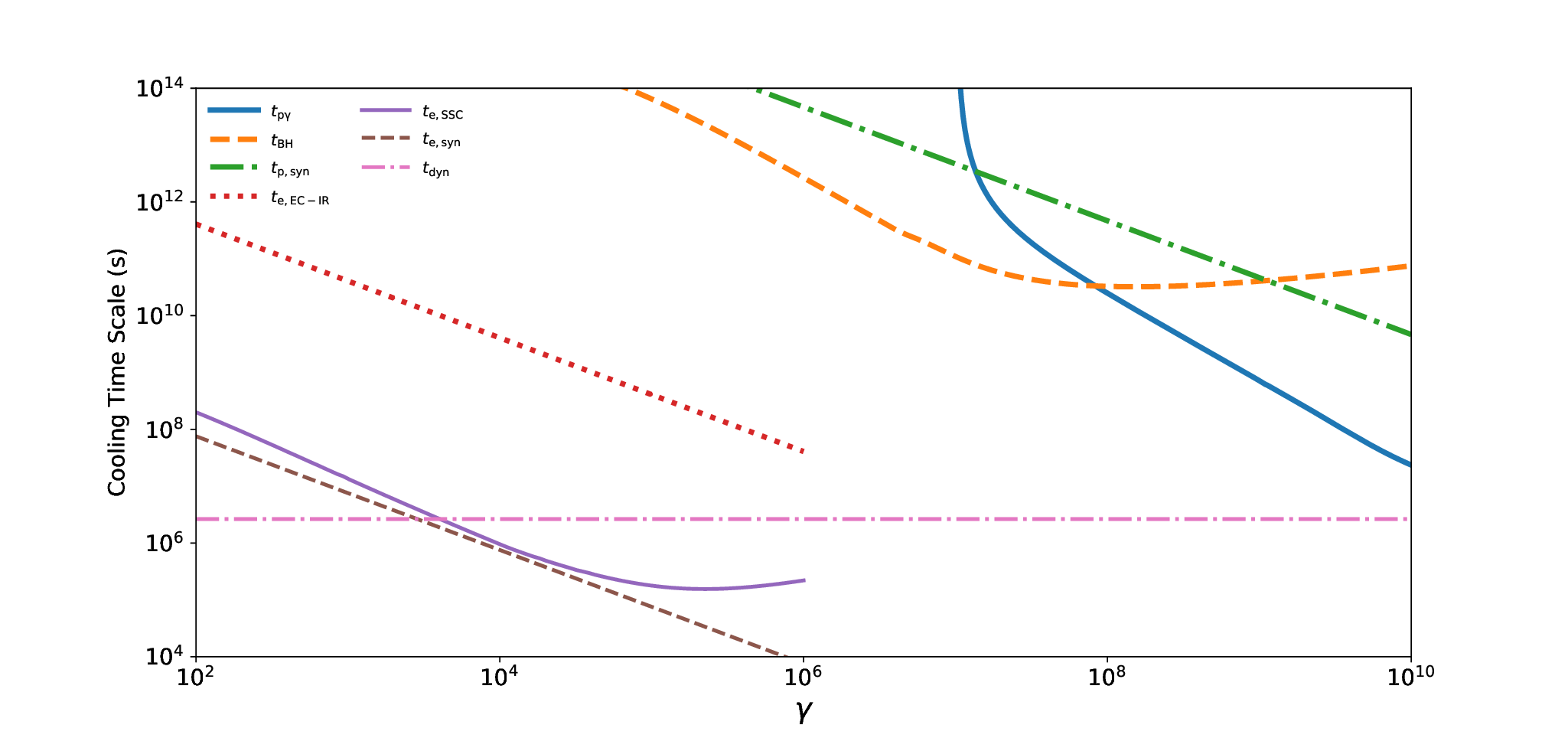}
   \caption{Timescales of various cooling processes for electrons and protons as a function of particle energy during the epoch 60260-60310\,MJD.
   The inset legend provides an explanation for all the curves.
   \label{fig:timescales.eps}}
\end{figure*}

\begin{figure*}
   \centering
   \includegraphics[width=1.0\textwidth]{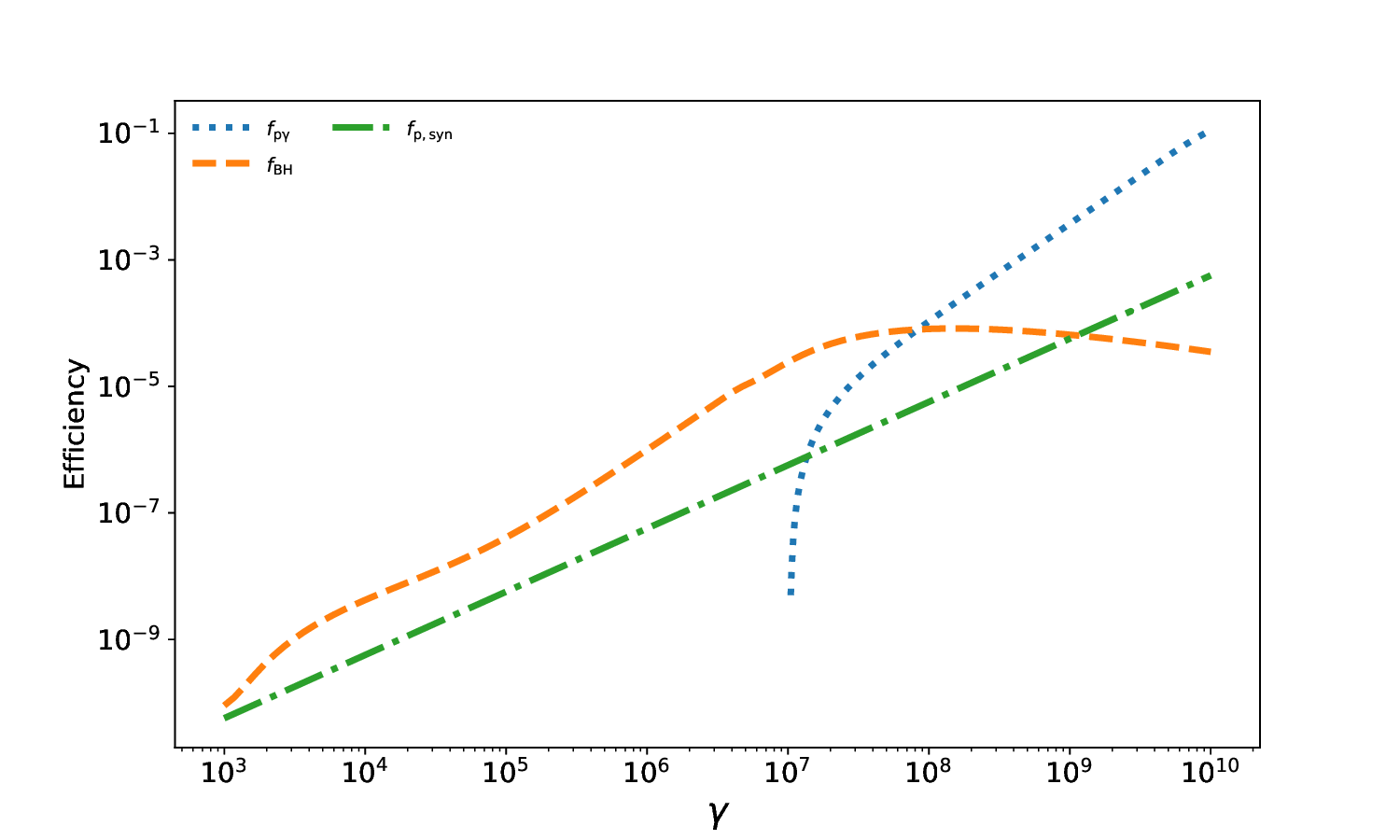}
   \caption{Efficiencies of various hadronic processes as a function of particle energy during the epoch 60260-60310\,MJD.
   The inset legend provides an explanation for all the curves.
   \label{fig:Efficiency.eps}}
\end{figure*}

The timescale for the Bethe Heitler (BH) process is given as 
\begin{equation}
\begin{split}
t_{BH}(\gamma_p) = 
\Bigg[\frac{7 \, m_e \, \alpha_f \, c \, \sigma_T}{9 \sqrt{2} \pi \, m_p \, \gamma_p^2 \,} 
\int_{\frac{1}{\gamma_p}}^{\infty} \frac{n_{\rm ph}(\epsilon)}{\epsilon^2}\mathrm{d}\epsilon \; \\
\left[ 
\left( 2 \gamma_p \epsilon \right)^{\frac{3}{2}} 
\left( \ln \left( \frac{2 \gamma_p \epsilon}{K_{BH}} \right) - \frac{2}{3} \right) 
+ \frac{2}{3} K_{BH}^{\frac{3}{2}}
\right] \Bigg]^{-1}.
\end{split}
\end{equation}
Here, $\alpha_f$ is the fine structure constant, and $K_{BH}$ is an adjustable constant $2\lesssim K_{BH} \lesssim 6.7$ (which is chosen as 3 in this work) \citep{Dermer-2009}.
The cooling timescales of various processes for electrons and protons as a function of the particle Lorentz factor during the VHE flaring epoch (60260-60310\,MJD) are shown in Figure $\ref{fig:timescales.eps}$. The energy loss timescales for leptonic processes are higher for EC than SSC and synchrotron processes, but for p$\gamma$ and BH processes, the target photon field is only the synchrotron radiation of primary electrons, which leads to the much longer cooling timescales of protons. These results are consistent with the previous studies \citep{Xue-2019, Xue-2023}.
From Figure $\ref{fig:timescales.eps}$, it is clear that the synchrotron cooling timescale is shortest, followed by the SSC and EC cooling timescales for electrons. Nevertheless, the cooling timescales for BH and photo-meson processes are longer; the proton synchrotron cooling timescale is the slowest (see Figure $\ref{fig:timescales.eps}$). The shortest cooling timescales of synchrotron emissions imply that electron cooling dominates the radiative processes in the system at lower energy, particularly up to the X-ray regime. On the other hand, the longer cooling times for photomeson and BH processes suggest that proton interactions contribute to high-energy photon production, primarily in the $\gamma$-ray range. Thus, synchrotron and SSC processes shape the optical/UV/X-ray to intermediate energy emissions, and the photomeson/BH processes dominate the $\gamma$-ray emission during the brightest $\gamma$-ray flare (60260-60310\,MJD). 
In the case of the photomeson process, the pions decay into $\gamma$-ray photons, contributing to the observed $\gamma$-ray emission, including the VHE emission. Additionally, there is also the possibility of secondary emission from the electromagnetic cascade. This secondary emission contributes to the broadband electromagnetic spectrum via synchrotron and inverse Compton (IC) processes. As highlighted by \citet{Keivani-2018, Xue-2019}, and \citet{Gao-2019} such cascades are expected to enhance the X-ray flux. However, in our observations, the X-ray flux enhancement during the flaring state was significantly lower compared to the $\gamma$-ray flux enhancement. This result suggests that cascade emission is not significant, and hence the $\gamma$-ray emission is primarily produced directly by the photomeson process. 
 This conclusion aligns with the findings of \citet{Gao-2019} and \citet{Xue-2019}, who also noted that a lack of significant X-ray emission could impose strong constraints on the efficiency of EM cascade in the broadband emission of blazars. To further investigate cascade emission from photomeson and BH processes, we calculated the efficiency of these mechanisms and the opacity to $\gamma-\gamma$ interaction. VHE $\gamma$-ray photons may interact with the low-energy photons, 
 for uniform isotropic photon fields, the opacity $\tau_{\gamma \gamma}$ is given by \citep{Dermer-2009, Xue-2023}
\begin{eqnarray}
\tau_{\gamma \gamma}(\epsilon_1) = \frac{R \pi r_{\rm e}^2}{\epsilon_1^2}
\int^\infty_{1/\epsilon_1} d\epsilon\ n_{\rm soft}(\epsilon)\ \bar{\phi}(s_0)\epsilon^{-2},\ 
\end{eqnarray}
where $\epsilon$ and $\epsilon_1$ are the dimensionless energies of low-energy and high-energy photons, $n_{\rm soft}$ is the number density of soft photons, $s_0=\epsilon \epsilon_1$. The term $\bar{\phi}(s_0)$ is expressed as
\begin{eqnarray}
\bar{\phi}(s_0) = \frac{1+\beta_0^2}{1-\beta_0^2}\ln w_0 - 
\beta_0^2\ln w_0 - \frac{4\beta_0}{1 - \beta_0^2}
\\ \nonumber 
+ 2\beta_0 + 4\ln w_0 \ln(1+w_0) - 4L(w_0)\ ,
\end{eqnarray}
with
$\beta_0^2 = 1 - 1/s_0$, $w_0=(1+\beta_0)/(1-\beta_0)$, and 
\begin{eqnarray}
L(w_0) = \int^{w_0}_1 dw\ w^{-1}\ln(1+w)\ .
\end{eqnarray}
For a synchrotron photon field, our results (Figure \ref{fig:Efficiency.eps}) show that VHE photons escape as the opacity is less than unity. This indicates that low-energy photons have minimal impact on the VHE $\gamma$-rays emitted from the region, consistent with \citet{Xue-2022}. A high optical depth is important for initiating pair cascades, and our results suggest a low likelihood of cascade development for VHE photons.
In our broadband SED model, we assumed IR photons at temperature 1000K as the target for the EC process. 
The Compactness parameter defining the pair production opacity can be expressed as $l^{'} = \frac{4\pi U^{'} \sigma_{T} R}{m_{e}c^2}$ \citep{Ghisellini-2009}, where $U^{'}$ is the external energy density, and $R$ is the size of the blob.  
For the selected IR target photon field, $l^{'}\sim 10^{-6}$ implies that VHE $\gamma$-rays are not absorbed within the blob, making EM cascades unlikely. To impose additional constraints on EM cascade emission, we also calculated the efficiency of hadronic processes. Using the dynamical timescale $t_{dyn} = \frac{R}{c}$ with the emission region size $R$ equal to $7.9\times 10^{16}$cm, one can calculate the efficiencies of hadronic processes like photo-meson as $\rm f_{p\gamma} = \frac{t_{dyn}}{t_{p\gamma}}$, BH process as $\rm f_{BH} = \frac{t_{dyn}}{t_{BH}}$, and proton-synchrotron process as $\rm f_{p,syn}=\frac{t_{dyn}}{t_{p,syn}}$. The efficiencies of these processes are plotted against the proton energy in Figure $\ref{fig:Efficiency.eps}$. The results (Figure \ref{fig:Efficiency.eps}) indicate that the efficiency of proton-synchrotron and BH process are inefficient, while the efficiency of $p\gamma$ interaction becomes significant at higher proton energies. 
Combining the opacity and efficiency results, our analysis supports the conclusion that $\gamma$-ray photons escape the emission region without contributing much to the EM cascade emission.
We calculated the proton jet power using $\rm P_{pr} = \pi R^2 \Gamma^2 \beta_{\Gamma} c U_{p}$ \citep{Celotti-2008, Gasparyan-2023}, where $\rm U_{p}$ is the proton energy density, $\rm \beta_{\Gamma}c$ is the velocity of emission region. For the chosen parameters we found $\rm P_{pr}=2.51\times 10^{43} erg s^{-1}$.\\
\section{Summary} \label{summary}
We performed a broadband spectral study of the VHE blazar B2\,1308+326 using data collected across various energy bands. The main motivation of the present work is to compare the broadband SED of the blazar during the epoch of VHE detection with the SED generated at different epochs. 
The summary of our study are
\begin{itemize}
\item The source was at its peak brightness in the $\gamma$-ray band during 60260-60310\,MJD. During this period, a VHE emission of the source was detected. This epoch was also associated with an enhanced flaring activity in the X-ray range.
\item Based on the SED modeling during the epochs 59250-59320\,MJD and 59750-59800\,MJD, it can be inferred that the observed $\gamma$-ray emission is a result of the interaction between relativistic jet electrons and thermal IR photons through IC scattering. However, during the epoch of VHE detection 60260-60310\,MJD, the $\gamma$-ray spectrum cannot be explained under leptonic emission models.
\item The inclusion of the photo-meson process along with the leptonic emission component can successfully reproduce the SED during 60260-60310\,MJD. This suggests that the source B2\,1308+326 has a non-negligible hadronic contribution during 60260-60310 MJD, but during the flaring period 59750-59800\,MJD, and quiescent period 59250-59320\,MJD, the leptonic process dominates over the hadronic process.\\
\item Further preliminary analysis of the LST-1 data reported by the LST team during December 11-14, 2023, shows an integrated flux above 100 GeV from the source as 15\% of the Crab Nebula. The detection is reported at a significance greater than $5$ sigma above 100 GeV. The constant VHE $\gamma$-ray flux from the Crab for the LST observation is reported as $\rm F_{>100 \rm GeV} = (4.95 \pm 0.03) \times 10^{-10} \rm cm^{-2} s^{-1}$\citep{Abe-2023}. In our analysis, 15\% of this flux was used to fit the broadband SED above 100 GeV. We noted that the lepto-hadronic model fits the broadband SED, including the VHE flux point. This result further supports our inference that during the VHE emission, a significant hadronic emission component is required to explain the broadband SED.\\
\item Moreover, based on the cooling timescales of various radiative processes, our study suggests that this excess $\gamma$-ray emission during the brightest $\gamma$-ray flare (60260-60310\,MJD) is likely due to the photo-meson process rather than EM cascades, as the X-ray flux enhancement was significantly lower than that of $\gamma$-rays.
\end{itemize}

\newpage



\begin{figure*}
   \centering
   \includegraphics[width=0.7\textwidth,angle=360]{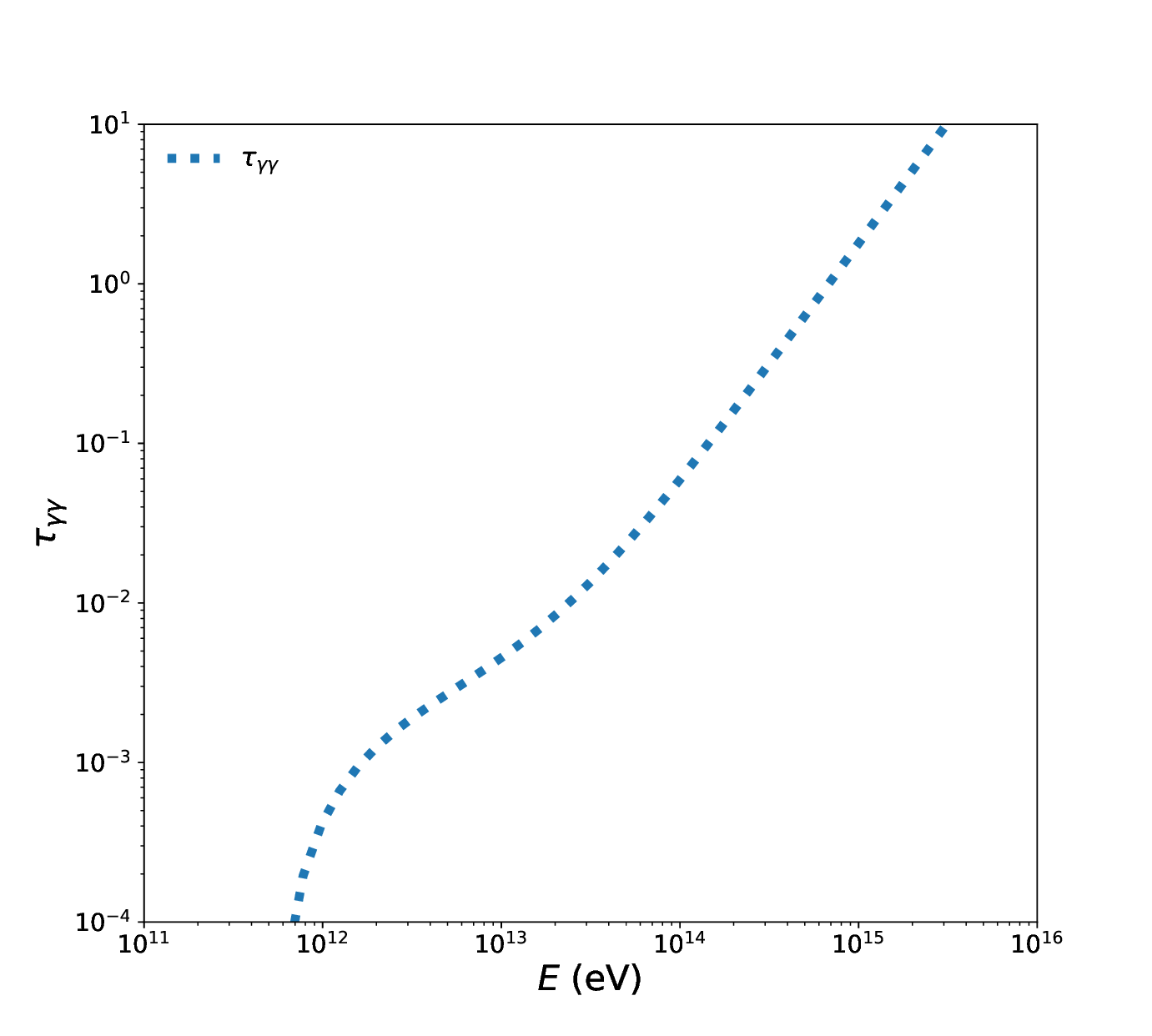}
   \vspace{0.5cm}
   \caption{The internal $\gamma\gamma$ opacity $\tau_{\gamma\gamma}$ as a function of high photon energy for B2\,1308+326 during the epoch 60260-60310\,MJD.
   \label{fig:opacity}}
\end{figure*}

    \begin{table*}[!htb]
    \centering
    \caption{\label{tab:sed_res}Best-fit model parameters for the broadband SEDs of B2\,1308+326 in the quiescent and flaring states. The size of the emission region ($R$) is fixed at $7.9\times 10^{16}$cm, and viewing angle ($\theta$) at $2^{\circ}$. The values in subscript and superscript for parameters in the model represent their lower and upper errors, respectively, obtained through the broadband spectral fitting. $--$ symbol indicates that the parameter's upper/lower error value is not constrained.}
    \begin{tabular}{lcccc} \hline
Free Parameter                       & Symbol &  State I  & State II  & State III \\ \hline
Low energy spectral index       & p                  		&   $1.90^{+0.38}_{-0.27}$                &$1.91^{+0.01}_{-0.04}$ & $1.87^{--}_{--}$ \\
High energy spectral index      & q                 	&   $4.04^{+0.13}_{-0.13}$                & $5.12^{+0.21}_{-0.15}$ & $4.03^{--}_{--}$ \\
Bulk Lorentz factor                     & $\Gamma$    & 
 $14.03^{+3.81}_{-1.70}$                 & $28.57^{+8.13}_{-6.13}$ & $24.12^{--}_{--}$ \\
Magnetic field                  & B   (G)                	&   $0.38^{+0.02}_{-0.02}$                & $0.38^{+0.01}_{-0.01}$ & $0.32^{--}_{--}$\\
\hline
Fixed Parameters\\
\hline
Low energy Lorentz factor  & $\gamma_{\rm min}$  & 14 &  14 & 150\\
High energy Lorentz factor & $\gamma_{\rm max}$$\times 10^{6}$ & 1& 1& 1\\
Break Lorentz factor & $\gamma_{\rm b}$ $\times 10^3$ & 1.6 & 6.4 & 2.5 \\ 
Energy density & $U_{*} \times 10^{-5}$ $(\rm erg cm^{-3})$ & 2.27 & 0.756 & 0.0756\\
\hline 
    \end{tabular}
    \label{tab:best-fit_parameters}
\end{table*}    
\section{Acknowledgments}
The authors thank the anonymous referee for valuable and insightful comments. which have greatly improved the quality of our manuscript.
ZS is supported by the Department of Science and Technology (DST), Govt. of India, under the INSPIRE Faculty grant (DST/INSPIRE/04/2020/002319). AD also thanks DST for providing financial support. AD is thankful to the Inter-University Centre for Astronomy and Astrophysics (IUCAA) Pune for hosting regular visits as part of its visitor program, during which a portion of this work was completed. NI acknowledges the IUCAA Pune, India, for support via associateship and hospitality. This research has used $\gamma$-ray data from the Fermi Science Support Center (FSSC). The work has also used the Swift Data from the High Energy Astrophysics Science Archive Research Center (HEASARC) at NASA's Goddard Space Flight Center.\\
\section*{Data Availability}
The data used in this paper are publicly available from the
archives at \url{https://heasarc.gsfc.nasa.gov/} and \url{https://Fermi.gsfc.nasa.gov/}. The models used in this work will be shared on
reasonable request to the corresponding author, Athar Dar (email: ather.dar6@gmail.com).
\bibliography{sample631}{}
\bibliographystyle{aasjournal}
\end{document}